\documentclass[sn-mathphys-num]{sn-jnl}

\usepackage{graphicx}%
\usepackage{multirow}%
\usepackage{amsmath,amssymb,amsfonts}%
\usepackage{amsthm}%
\usepackage{mathrsfs}%
\usepackage[title]{appendix}%
\usepackage{xcolor}%
\usepackage{textcomp}%
\usepackage{booktabs}%
\usepackage{algorithm}%
\usepackage{algorithmicx}%
\usepackage{algpseudocode}%
\usepackage{listings}%

\theoremstyle{thmstyleone}%
\theoremstyle{thmstyletwo}%

\theoremstyle{thmstylethree}%

\begin{document}

\title[Article Title]{Self-organized first-order transition from foreshock to mainshock in earthquake sequences induced by heat, fluid pressure, and porosity}


\author*[1]{\fnm{Takehito} \sur{Suzuki}}\email{t.suzuki@takachiho.ac.jp}

\author[2]{\fnm{Hiroshi} \sur{Matsukawa}}\email{matsu@phys.aoyama.ac.jp}


\affil*[1]{\orgdiv{Faculty of Human Sciences}, \orgname{Takachiho University}, \orgaddress{\street{2-19-1 Omiya}, \city{Suginami-ku}, \postcode{168-8508}, \state{Tokyo}, \country{Japan}}}

\affil[2]{\orgdiv{Department of Physical Sciences}, \orgname{Aoyama Gakuin University}, \orgaddress{\street{5-10-1 Fuchinobe, Chuo-ku}, \city{Sagamihara}, \postcode{252-5258}, \state{Kanagawa}, \country{Japan}}}


\abstract{
Earthquake cycles are studied 
by taking into account the interactions among slip, fluid pressure, temperature, and porosity on the fault planes, which are known to play a crucial role in earthquake dynamics.
The spring-block model with a single block is employed.
A first-order transition from foreshock to mainshock occurring spontaneously in earthquake sequences is discovered both analytically and numerically. 
This transition is induced by these interactions.
It is shown that the function of the slip distance $u$, $F(u)$, 
defined as the sum of the difference between the energies stored in the driving spring before and after the slippage, and the energy dissipated during the slippage, governs the transition.
The equation, $F(u)=0$, represents the energy balance before and after  the slippage, and the solution $u=u_f$ describes the realized slip distance for each slippage event.
The solutions discontinuously transition 
from small to large slippages in the sequence of earthquakes.
This transition can be interpreted to be a self-organized first-order transition 
from small to large slippages.
The former slippage is governed by pore generation, whereas the latter is governed by thermal pressurization.
A phase diagram of the foreshocks and mainshocks, which is also considered a phase diagram of slow and fast earthquakes, is obtained.
}




\maketitle

\section{Introduction} \label{secInt}


In natural fault zones, we sometimes observe that the foreshocks repeat several times and change to the mainshock. 
The slip velocities of foreshocks are typically on the several order of magnitude smaller than those of mainshocks.
The transition time from foreshocks to the mainshock and the transition mechanism remain controversial.

The foreshock-mainshock behavior corresponds to those of precursors inducing macroscopic slippage. 
When shear stress is applied to a block on a substrate, it is believed that the block moves as a whole only when the shear stress exceeds the macroscopic static friction stress.
However, in laboratory experiments, local slip events with slip distances and velocities much smaller than those of macroscopic slippage occur below the macroscopic static friction stress \cite{Rub}.
These precursors have been investigated using the visco-elastic model with the local slip-velocity weakening friction law introduced phenomenologically  in numerical and theoretical studies \cite{Ots}. 
In this case, the lengths of the precursor propagation determine the transition to the macroscopic slippage \cite{Rub, Ots, Kata14}. 


The interactions among the slip,
heat, fluid pressure, and porosity are known to play a crucial role in a variety of dynamic earthquake slippage behavior \cite{Suz14, Suz17}. 
During the dynamic slippage, frictional heating occurs and pores are generated due to the fracture of fault rocks.
The former [latter] increases [decreases] the fluid pressure, inducing the reduction [increase] in the friction stress and an increase [decrease] in the slip velocity.  
This feedback has explained several aspects of dynamic earthquake slippage behaviors, such as the slip-pulse propagation \cite{Suz08} and the generation mechanism of slow earthquakes \cite{Suz09, Seg}. 
The increase in the fluid pressure due to frictional heating is referred to as thermal pressurization.

Earthquake sequences have been widely investigated in seismology (e.g., \cite{Kin, Ben, McG}). 
Most of those works, however,  do not take into account the effects of the important interactions among the slip,
heat, fluid pressure, and porosity, mentioned above.
Zilio \textit{et al.} investigated earthquake sequences by taking into account 
the effects of slip, fluid pressure, and porosity, but not heat \cite{Zil}.
Systematic understanding from the viewpoint of all of these effects has, however, not been performed.

In this study, we analyze the sequence of slippages by considering the interactions among slip, heat, fluid pressure, and porosity, based on the spring--block model which has been widely employed to simulate seismicity \cite{Car}.
It is shown that the slip is inhibited and promoted by the pore-generation and thermal-pressurization effects, respectively. 
The inhibited and promoted slippages are referred to here as small and large slippages, respectively.
Let us consider that a small slippage occurs initially
after driving the upper substrate. 
If the driving continues, the small slippages repeat several times. 
After each small slippage, a slip deficit, defined as the difference between the extension length of the spring just before the dynamic slippage and each slip distance, appears.
The slip deficit accumulates with the repetition of the small slippages.
Eventually, the small slippages end and transition into large slippages
compensate for the slip deficit. 
The dominant mechanism in the slip process  changes from pore generation to thermal pressurization. 
This small-large slippage transition in the earthquake sequence corresponds to the foreshock-mainshock transition, 
or slow to fast earthquake transition 
from a seismological viewpoint. 
It is to be emphasized that the foreshock-mainshock transition investigated in this work occurs spontaneously in a single framework  without any fitting parameters due to the interaction among slip, heat, fluid pressure and porosity.

The first-order phase transition between the creep-like motion and the stick-slip motion on the spring-block model has been reported \cite{deS, Vas}. 
The transition investigated in these studies occurs by varying the velocity scale appearing in the friction law.
These studies shed light on the important role of the velocity scale.
The transition discussed in these studies does not occur, however, in the sequence of earthquake in a single system with the same friction parameter.
The transition from the foreshock to main shock with recurrence of earthquakes should be explained in a single framework.

This paper is organized as follows. The model setup is clarified in Sec.~\ref{secMS}, and the details of the spring-block model and the interaction among the slip, heat, fluid pressure, and porosity are described therein. 
The analytical treatment of the model is described in Sec.~\ref{secAT}. 
The analytical treatment reveals that the transition from foreshocks to mainshocks can be considered a self-organized first-order phase transition caused by the repetition of the foreshocks.
The analytical result is visualized by numerical calculations, as described in Sec.~\ref{secNC}. 
The numerical calculations are consistent with the phase diagram of the small and large slippages in the 
pressure -- porosity plane at the onset of slippage
which is analytically obtained.
This study's findings are summarized and discussed in Sec.~\ref{secDisCon}.

\section{MODEL SETUP} \label{secMS}

\begin{figure}[tbp]
\centering
\includegraphics[width=8.5cm]{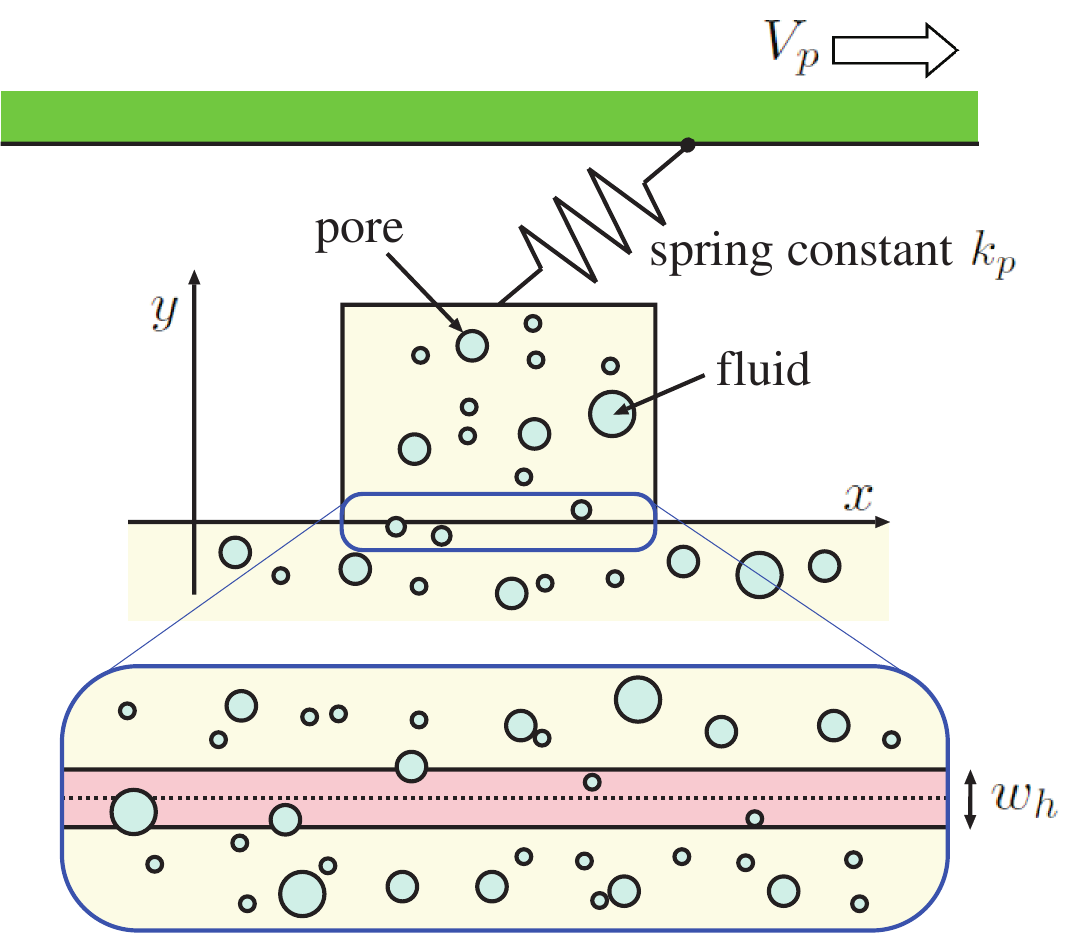}
\caption{Model setup: The block and the substrates represent the fault.
The block is driven by the upper plate via a spring.
The block and the substrate contain pores and fluid.  
The motion of the block generates frictional heat and induces pore formation, which alter the fluid pressure and feedback into the frictional properties, as explained in Sec. \ref{secIHFP} and \ref{secBK-IHFP}}
\label{FigMS}
\end{figure}

The spring-block model with the interaction among the slip, heat, fluid pressure, and porosity effects are employed to analyze the  transition from foreshocks to the mainshock.
The upper substrate moves with a constant velocity.
The model setup is illustrated in Fig.~\ref{FigMS}.
In this study, fluid in the pore in the block and the substrate and the frictional heating are taken into account as explained in the following subsections.
The details of the spring-block model are summarized in Sec.~\ref{secBM} and the interaction is described in Sec.~\ref{secIHFP}.
The implementation of the interaction in the spring-block model is provided in \ref{secBK-IHFP}.

\subsection{The spring-block model}  \label{secBM}

We employ a spring-block model consisting of a single block on a substrate, an upper plate, and a spring connecting them.
The motion of the upper spring increases the driving force acting on the block and stores elastic energy. 
When the driving force on a block exceeds the maximum static friction force, the block slips.
The macroscopic static friction force is given by the multiplication of the normal force acting on the contact area and the static friction coefficient. During the dynamic slippage, the stored elastic energy is released, and the dynamic friction force between the block and the lower substrate reduces the kinetic energy of the block. 
Consequently, the dynamic slippage eventually ceases. 
After the cessation, the driving force and stored elastic energy are again increased by the motion of the upper substrate, following which the next slippage occurs. 
This cycle of slippage and cessation is widely known as the stick-slip behavior, which corresponds to earthquake cycles \cite{Car}. 
Here, we examined the single block model and qualitatively analyzed the behavior of the earthquake cycle.

\subsection{Interaction among slip, heat, fluid pressure, and porosity} \label{secIHFP}

Fault rocks are considered to be poroelastic media, consisting of a solid skeleton and pores within it. 
The pores are usually considered to be filled with fluid such as water or oil. 
Let us consider that the dynamic earthquake slippage (frictional slippage) occurs inside the media. 
The slip is accompanied by macroscopic deformation in the zone around the slip interface. 
This zone is referred to as the slip zone below, and it is considered to have a finite width. 
Two kinds of fluid-pressure behaviors are considered inside the slip zone by addressing the interaction among the 
slip, 
heat, 
fluid pressure, and porosity \cite{Suz17}. 
When the temperature increases due to frictional heating, both the solid skeleton and the fluid expand. 
Since the thermal expansion of the fluid is larger than that of  the solid skeleton, the fluid pressure increases. 
This increase in the fluid pressure due to frictional heating is termed thermal pressurization. 
Conversely, if the total volume of pores increases inside the slip zone, which is called pore generation, and its effect is dominant, the fluid pressure decreases. 
Therefore, the interaction among the slip, heat, fluid pressure, and porosity effects can explain both the increase and decrease in the fluid pressure.

The interaction among the slip, heat, fluid pressure, and porosity  plays an important role in the dynamic earthquake slippage process \cite{Suz14, Suz17}. 
If the fluid pressure increases [decreases], the effective normal stress on the slip zone, i.e., the difference between the applied normal stress and the fluid pressure, decreases [increases]. 
The dynamic friction stress for porous media is given by the effective normal stress at the slip interface times the sliding friction coefficient.
Therefore, the dynamic friction stress also decreases [increases], inducing slip acceleration [deceleration]. 
The thermal pressurization favors the slip, i.e., inducing slip acceleration, whereas the pore generation inhibits the slip, i.e., inducing the slip deceleration. 
Therefore, the slip distance dominated by
the former  is significantly larger than that 
by the latter.
The acceleration due to the thermal-pressurization effect eventually ends because of the decrease of driving force via the spring.
Notably, the situation is different for the infinitely long mode III crack \cite{Suz17}, where constant shear stress works infinitely far from the crack and the slip velocity is proportional to the stress drop \cite{Bru}.
Also in that case, the acceleration due to the thermal-pressurization effect eventually ends.
This is because the fluid pressure approaches the normal stress and the effective normal stress approaches zero, eliminating the friction stress following the vanishing acceleration rate.

\subsection{Combining the spring-block model with the interaction among slip, heat, fluid pressure and porosity} \label{secBK-IHFP}

To address the transition from foreshocks to the mainshock, we employ the spring-block model with a single block. 
The block is considered a poroelastic medium, and we set it on a poroelastic substrate. 
Let us denote the mass of the block as $m$ and the base area as $S$.
The upper substrate is moving with constant velocity $V_p$ and drives the block via the spring with spring constant $k_p$. 
We consider the interaction among the slip, heat, fluid pressure, and porosity within the slip zone around the slip interface.
The framework constructed here enables us to address the earthquake sequence by taking into account those interactions, which have not been investigated in previous studies.

The frictional interface is set to be in the $x-z$ plane, and the loading direction is along the $x$ axis.
The variation along the $z$ direction is ignored.
The slip zone is assumed to locate in the region $w_h/2 \le y \le w_h/2 $, where $w_h$ is the thickness of the slip zone. 
The interaction among the slip, heat, fluid pressure and porosity effect emerges within this zone. 
We consider the fluid flow from/into the slip zone. The fluid flow is proportional to the gradient of the fluid pressure, and the occurrence of inflow or outflow is determined by the condition at the boundary of the slip zone, $y=\pm w_h/2$. 
If the pressure just inside the boundary is less [more] than that just outside, the fluid inflow [outflow] into [from] the slip zone occurs.
The fluid flow changes the effective normal stress in the slip zone, inducing the feedback in the slippage process and changing the timing of the onset and the end of the slippage.
%
%
We also consider the pore-healing effect during intervals between successive earthquakes \cite{Bye, Ren, Ten, Im}.
The quantitative framework addressing pore healing is explored in the following subsection.

\subsection{Governing Equations} \label{secGE}

As noted in Sec. \ref{secBM}, the block repeats the stick-slip behavior. 
We consider the governing equations during the stick time and dynamic slippage. 
First, the governing equations during the dynamic slippage are \cite{Suz17}
\begin{equation}
\frac{1}{M_0} \frac{\partial p}{\partial t} =
\tilde{\alpha}\frac{\partial T}{\partial t} +\frac{k}{\eta} \frac{\partial^2 p}{\partial y^2} -\frac{\partial \phi}{\partial t}, \label{eqDPEL}
\end{equation}
\begin{eqnarray}
& &
\tilde{C}\frac{\partial T}{\partial t} 
=
 \begin{cases}
\dfrac{\sigma_{\mathrm{res}} v} {w_h}  &  w_h/2 \le y \le w_h/2, \\
0 & \text{ otherwise},
\end{cases}
 \label{eqDTEL}
\end{eqnarray}
\begin{equation}
\frac{\partial \phi}{\partial t}=
\begin{cases}
\alpha_0 v \left( 1-\dfrac{\phi}{\phi_\infty} \right)  &  w_h/2 \le y \le w_h/2, \\
0 & \text{ otherwise},
\end{cases}
\label{eqDPhiEL}
\end{equation}
\begin{equation}
m \frac{d^2 u}{dt^2} =k_p (V_p t -u) +\mu_{\mathrm{slid}} (\sigma_n^0 +p)S, \label{eqeom}
\end{equation}
where $p$ fluid pressure, $T$ temperature, 
$\phi$ the porosity increment from the reference value $\phi_{\mathrm{ref}}$ which is the porosity at $t=0$, 
$u$ the slip distance of the block, $v=du/dt$ the slip velocity, $M_0$ the effective bulk modulus, $\tilde{\alpha}$  effective thermal expansion coefficient, and $\tilde{C}$ effective heat capacity.
Refer to Table 1 for the definitions of the parameters.

Equations (\ref{eqDPEL})--(\ref{eqDPhiEL}) describe the temporal evolutions of $p$, $T$, and $\phi$ during the dynamic slippage, respectively. 
The terms on the right-hand side of Eq.~(\ref{eqDPEL}) describe the thermal-pressurization, fluid-diffusion, and pore-generation effects, respectively. 
The term on the right-hand side of Eq.~(\ref{eqDTEL}) describes the heat source term, i.e., the residual friction stress times the slip velocity divided by $w_h$. 
The porosity evolution law (\ref{eqDPhiEL}) describes the monotonic increase in $\phi$ with increasing slip distance to the upper limit $\phi_{\infty}$ \cite{Suz14, Suz17}. 
Equation (\ref{eqeom}) is the equation of motion of the block. 
The first and second terms on the right-hand side of Eq.~(\ref{eqeom}) show the driving stress from the upper spring and the friction stress, respectively. 
Notably, the normal stress is defined as negative, $\sigma_n^0 \le 0$, and the effective normal stress $\sigma_n^0 +p_0 \le 0$\ holds anytime.


We then show the evolution laws during the stick time. 
During the stick time, $u$ is 
constant.
We write the value of $\phi$ at each slip cessation, which is the initial value of the stick time, as $\phi_{\mathrm{cease}}$. During the stick time, we use the following equation system 
\begin{eqnarray}
& &
\tilde{C}\frac{\partial T}{\partial t} = 
\tilde{\lambda}\frac{\partial^2 T}{\partial y^2}, \label{eqSTEL}
\end{eqnarray}
\begin{equation}
\phi=\frac{\phi_\mathrm{cease}}{1+\alpha_1 t'}, \label{eqSPhiEL}
\end{equation}
where 
$\tilde{\lambda}$ is the effective thermal diffusion constant,
$\alpha_1$ is a positive constant, and $t'$ represents the time measured from the slip cessation. Equations (\ref{eqSTEL}) and (\ref{eqSPhiEL}) describe the temporal evolutions of $T$ and $\phi$, respectively. 
As noted in Sec.~\ref{secBK-IHFP}, $\phi$ decreases from $\phi_{\mathrm{cease}}$ during the interval between successive earthquakes owing to pore healing \cite{Ren}. 
Though it is difficult to express the exact porosity evolution law in a simple manner \cite{Ren}, we employ Eq.~(\ref{eqSPhiEL}) as an approximate form. Finally, the temporal evolution equation of $p$ during the stick time is the same as that during the slip time, Eq. (\ref{eqDPEL}).

\section{Analytical treatment} \label{secAT}


\subsection{Function governing the  transition from foreshocks to the mainshock} \label{secF}

Here, we derive the function governing the transition from foreshocks to the mainshock in terms of the slip distance for each slippage event. 
We consider the governing equations during the dynamic slippage. 
First, we neglect the fluid-diffusion term, $(k/\eta) \partial^2 p/\partial y^2$, in Eq.~(\ref{eqDPEL}) during the dynamic slippage for the analytical treatment. 
When the permeability is smaller than approximately $10^{-18} \ [\mathrm{m^2}]$, this approximation is verified \cite{Suz10}. 
With this approximation, the slip zone becomes an isolated system, and variables do not depend on $y$,
i.e., all the variables are functions of only time and are independent of position. 
Additionally, we neglect the dependence of coefficients appearing in Eqs.~(\ref{eqDPEL}) and (\ref{eqDTEL}) on $\phi$ (these coefficients will be referred to as COEs hereinafter). 
The validity of this approximation is discussed in the next section.
Therefore, the governing equations, i.e., (\ref{eqDPEL})--(\ref{eqDPhiEL}) in the slip zone during the dynamic slippage are given by
\begin{equation}
\frac{dp}{dt}=C_1 \frac{dT}{dt} -M_0 \frac{d \phi}{dt}, \label{eqGP}
\end{equation}
\begin{equation}
\frac{dT}{dt}=-C_2 (\sigma_n^0+p) \frac{du}{dt}, \label{eqGT}
\end{equation}
\begin{equation}
\frac{d \phi}{dt}=\alpha_0 \frac{du}{dt} \left( 1-\frac{\phi}{\phi_\infty} \right), \label{eqGPhi}
\end{equation}
respectively, where $C_1 \equiv M_0 ((b-\phi_{\mathrm{ref}}) \alpha_s +\phi_{\mathrm{ref}} \alpha_f)$ and $C_2 \equiv \mu_{\mathrm{slid}}/w_h((1-\phi_{\mathrm{ref}}) \rho_s C_s +\phi_{\mathrm{ref}} \rho_f C_f)$ are positive constants. 

Using Eqs.~(\ref{eqGP})--(\ref{eqGPhi}), we derive the function governing the transition from foreshocks to the mainshock. 
Equation (\ref{eqGPhi}) gives $\phi$ as a function of  $u$ as follows:
\begin{equation}
\phi=\phi_\infty-(\phi_\infty-\phi_0) e^{-\alpha_0 u/\phi_\infty}, \label{eqPhiU}
\end{equation}
where $\phi_0$ is the value of $\phi$ at the onset of each dynamic slippage. 
The time evolution of $p$ is obtained from Eqs.~(\ref{eqGP})--(\ref{eqPhiU}) as follows:
\begin{equation}
\frac{dp}{dt}=\left( -\gamma (\sigma_n^0+p) -M_0 \left( 1-\frac{\phi_0}{\phi_\infty} \right) \alpha_0 e^{-\alpha_0 u/\phi_\infty} \right) \frac{du}{dt},
\end{equation}
where $\gamma =C_1 C_2$. 
Thus, we can express $p$ in terms of $u$ as,
\begin{equation}
p=\left( \sigma_n^0+p_0 -A  \right)e^{-\gamma u}+A e^{-\gamma' u} -\sigma_n^0, \label{eqPU}
\end{equation}
where $A \equiv M_0 (1-\phi_0/\phi_{\infty})  \alpha_0/(\gamma'-\gamma), \gamma' \equiv \alpha_0/\phi_\infty$, and $p_0$ is the fluid pressure at the instant of each dynamic slippage onset.

We now consider the energy balance of the block and the upper spring system
just before and after the slippage. 
Just before the slippage, the loading stress from the upper spring must be equal to the maximum static friction stress.
Here is the expression:
\begin{equation}
k_p L =-\mu_{\mathrm{stat}} (\sigma_n^0+p_0) S, \label{eqMSF}
\end{equation}
where $L$ is the extension of the upper spring just before the slippage, and $\mu_{\mathrm{stat}}$ is the static friction coefficient.
The energy stored in the upper spring at that instance, $E_1$, is given by 
\begin{equation}
E_1=\frac{1}{2}k_p L^2=\frac{1}{2} \frac{\mu_{\mathrm{stat}}^2}{k_p} (\sigma_n^0+p_0)^2 S^2. \label{eqE1}
\end{equation}
Next, we consider the energy stored in the upper spring
at the slip distance $u$,  $E_2$. 
It is reasonable to neglect the motion of the upper substrate during the slippage. 
With this approximation, Eq.~(\ref{eqMSF}) gives the expression of $E_2$ as follows:
\begin{equation}
E_2=\frac{1}{2} k_p (L-u)^2=\frac{1}{2} k_p \left(-\frac{\mu_{\mathrm{stat}}}{k_p} (\sigma_n^0+p_0) S -u \right)^2 .
\label{eqE2}
\end{equation}
The work performed on the block during the dynamic slippage by the friction force is obtained using Eq.~(\ref{eqPU}):
\begin{eqnarray}
E_{\mathrm{fric}}=\int_0^{u} \left(-\mu_{\mathrm{slid}} (\sigma_n^0+p) \right) S du' \nonumber \\
=-\mu_{\mathrm{slid}} S \int_0^{u} \left[ \left( \sigma_n^0+p_0 - A  \right)e^{-\gamma u'} \right. \nonumber \\
\left. + A e^{-\gamma' u' } \right] du' \nonumber \\
=-\mu_{\mathrm{slid}} S \left[ \frac{1}{\gamma} \left( \sigma_n^0+p_0 -A \right) (1-e^{-\gamma u}) \right. \nonumber \\
\left. + \frac{1}{\gamma'} A (1-e^{-\gamma' u }) \right]. \label{eqEf}
\end{eqnarray}
The lower bound for the integration is not 
$-L$ 
but zero because $u$ describes the slip increment within a single event. 

The energy change
after the slip with the distance $u$, in addition to the work done by the kinetic friction force
is expressed by the function $F(u)$,
\begin{equation}
F(u)=E_2-E_1+E_{\mathrm{fric}}. \label{eqEB}
\end{equation}
By employing Eqs.~(\ref{eqE1})--(\ref{eqEf}), we obtain
\begin{align}
F(u) &=\frac{1}{2} k_p u^2 +\mu_{\mathrm{stat}}(\sigma_n^0+p_0) S u \nonumber \nonumber\\
&-\mu_{\mathrm{slid}} \left[ \frac{1}{\gamma} \left( \sigma_n^0+p_0 -A  \right) (1-e^{-\gamma u}) \right. \nonumber \\
&\left. + \frac{1}{\gamma'} A (1-e^{-\gamma' u }) \right] S. 
\label{eqCrit}
\end{align}
The energy balance before and after each slippage event with the slip distance $u_f$ is expressed as
\begin{eqnarray}
F(u = u_f)&=& E_2-E_1+E_{\mathrm{fric}} \nonumber\\
&=& 0 \label{eqCrit0}
\end{eqnarray}
This is the equation that $u_f$ satisfies. Actually, the interaction among slip, heat, fluid pressure, and porosity is a non-equilibrium process with dissipation because of the friction stress.
In many works of the non-equilibrium systems with dissipation, the variational principle with the Rayleigh function is employed \cite{Doi, Fuk}.
In this section, we are interested in the slip distance of each slippage event, not in the behavior during slippage; the employment of the function of $F(u)$ and the condition of $F(u_f)=0$ enhance our perspective on the issue at hand. The equation of motion is not used for this treatment.

The behavior of the solutions for Eq.~(\ref{eqCrit0}) determines the  transition from foreshocks to the mainshock. 
The transition is considered a first-order phase transition as mentioned below. 

To investigate the solutions for Eq.~(\ref{eqCrit0}), we consider the boundary conditions for $F(u)$. 
The derivative of $F(u)$ is given as follows:
\begin{eqnarray}
\frac{d F(u)}{d u} &=& k_p u +\mu_{\mathrm{stat}}(\sigma_n^0+p_0) S \nonumber \\
&-&\mu_{\mathrm{slid}} \left[ \left( \sigma_n^0+p_0 -A \right) e^{-\gamma u} \right. \nonumber \\
&+& \left. A e^{-\gamma' u} \right] S. \label{eqDiffCrit1}
\end{eqnarray} 
From the forms of $F(u)$ and $dF(u)/du$, we can derive that $F(0)=0$ and $d F(u)/d u |_{u=0}<0$. 
We can also confirm that $\lim_{u \to \infty} F(u) = \infty$ and $\lim_{u \to \infty} dF(u)/du = \infty$. 
From Eq.~(\ref{eqDiffCrit1}), we also have
\begin{eqnarray}
\frac{d^2 F(u)}{d u^2}=k_p -\mu_{\mathrm{slid}} [ -\gamma (\sigma_n^0+p_0-A)e^{-\gamma u} \nonumber \\
-\gamma' A e^{-\gamma' u} ] S, \label{eqDiffCrit2}
\end{eqnarray}
and
\begin{eqnarray}
\frac{d^3 F(u)}{d u^3}=-\mu_{\mathrm{slid}} [ \gamma^2 (\sigma_n^0+p_0-A)e^{-\gamma u} \nonumber \\
+\gamma'^2 A e^{-\gamma' u} ] S. \label{eqDiffCrit3}
\end{eqnarray}
From Eq.~(\ref{eqDiffCrit3}), we  obtain the single  solution of $d^3 F(u)/du^3=0$ as
\begin{equation}
u_{\mathrm{sol}}=\frac{1}{\gamma-\gamma'} \ln \left( -\left( \frac{\gamma}{\gamma'} \right)^2 \frac{\sigma_n^0+p_0-A}{A} \right).
\end{equation}
For 
$u_{\mathrm{sol}}>0$ to be met, we need
\begin{equation}
-\left( \frac{\gamma}{\gamma'} \right)^2 \frac{\sigma_n^0+p_0-A}{A}>0  \label{eqC-Usol1}
\end{equation}
and
\begin{equation}
\frac{1}{\gamma-\gamma'} \left( -\left( \frac{\gamma}{\gamma'} \right)^2 \frac{\sigma_n^0+p_0-A}{A} -1 \right)>0. \label{eqC-Usol2}
\end{equation}
If conditions (\ref{eqC-Usol1}) and  (\ref{eqC-Usol2}) are satisfied, $d^2 F(u)/du^2=0$ will have at most two positive solutions and $d F(u)/du=0$ will have at most three positive solutions. 
Additionally, from the boundary conditions, $dF(u)/du|_{u=0}<0$ and $\lim_{u \to \infty} dF(u)/du =\infty$, equation $d F(u)/du=0$ has at least a single positive solution. 
Thus, the number of solutions to $dF(u)/du=0$ is one, two, or three. 
A physically interesting behavior is observed when three positive solutions exist.
Here we consider that case.
If the equation has two solutions, it is considered that the two of three solutions are degenerate.


We present three positive solutions for $dF(u)/du=0$, $u_1<u_2<u_3$. 
With the boundary conditions, $F(0)=0$ and $dF(u)/du|_{u=0}<0$, $F(u_1)$ must be negative. 
Therefore, we must consider three cases: (i) $F(u_2)>0$ and $F(u_3)>0$, (ii) $F(u_2)>0$ and $F(u_3)<0$, and (iii) $F(u_2)<0$ and $F(u_3)<0$. 
We first consider case (ii). 
In this case,  Eq.~(\ref{eqCrit0}), $F(u_f) =0$, has three positive solutions expressed as $u_{\mathrm{small}} < u_{\mathrm{middle}} < u_{\mathrm{large}}$. 
Among them, $u_{\mathrm{small}}$ is physically realized because $u_f$ starts from 0.
If $F(u_3)$ becomes positive, as is in case (i), $u_{\mathrm{middle}}$ and $u_{\mathrm{large}}$ disappear and Eq.~(\ref{eqCrit0}) has a single positive solution, which corresponds to $u_{\mathrm{small}}$.
The solution for case (i) is, however, not realized in our calculation and not considered henceforth.
If $F(u_2)$ becomes negative, as is in case (iii), $u_{\mathrm{small}}$ and $u_{\mathrm{middle}}$ disappear and  Eq.~(\ref{eqCrit0}) has only a single positive solution corresponding to $u_{\mathrm{large}}$. 
Thus, we can conclude that the realized solution of Eq.~(\ref{eqCrit}) discontinuously transitions from $u_{\mathrm{small}}$ to $u_{\mathrm{large}}$ at the point where $F(u_2)=0$ between cases (ii) and (iii).
This transition of the solution is interpreted as a first-order phase transition, where the order parameter is $u_f$, although the behavior is not a statistical mechanical one. 
Notably, the occurrence of the discontinuous transition of the solution 
depends on whether $F(u_2)$ is positive or negative. 




The physical interpretations of $u_{\mathrm{small}}$ and $u_{\mathrm{large}}$ are given as follows. 
There are two forms of slippages.
In the first form, the pore-generation effect dominates and induces the fluid-pressure decrease and friction-stress increase during the slippage. 
In this case, $u_f$ is significantly smaller than  the extension of the upper spring just before the slippage, $L$, and only a tiny fraction  of the strain stored in the spring is released.
This is the small slippage noted in Sec.~\ref{secInt}, and it corresponds to $u_{\mathrm{small}}$. 
In the other form, the thermal-pressurization effect dominates and induces the fluid-pressure increase and friction-stress decrease during the slippage. 
In this case, $u_f$ is close to $L$, and most of the energy stored in the spring is released after the slippage. 
This is the large slippage noted in Sec.~\ref{secInt}, and it corresponds to $u_{\mathrm{large}}$.
Slippages that appear in the sequence of repeating foreshocks and the mainshocks obtained by the numerical calculation in Sec.~\ref{secNC} are classified either as  $u_{\mathrm{small}}$ or $u_{\mathrm{large}}$.
We discuss only  $u_{\mathrm{small}}$ and $u_{\mathrm{large}}$. 

Note again that the transition between the small and large slippages is interpreted as the phase transition; these slippages are qualitatively different and discontinuous. 
However, both of these slippages can occur in a temporal sequence of slippage repetition, as follows. 
Notably, the $F(u)$ function includes $p_0$ and $\phi_0$.
The $p_0$-dependence of $F(u)$ is clearly indicated in Eq.~(\ref{eqCrit}), and the $\phi_0$-dependence emerges via $A$. 
We can insist that the occurrence of small or large slippages is strictly determined by the condition at the onset of each slippage. 
In the repetition of slippages, $p_0$ and $\phi_0$ can differ among slippages. 
The repetition of the small slippages inevitably results in large slippages, attributed to the accumulated slip deficit. 
We numerically show this repetition behavior in Sec.~\ref{secNC}.

The first-order transition associated with the spring-block model has been reported \cite{deS, Vas}.  Their transition was associated with the change in the friction law. The velocity-weakening friction law is assumed there, and the ratio of characteristic velocity for the weakening to that for the harmonic oscillation governs the transition.
The transition emerges when the parameters in the friction law are varied, and does not emerge by repeating slippages with the same friction law with the same parameters.

Numerical calculation plays another important role. 
As noted in the current section, we neglected the dependence of COEs on $\phi$. 
To confirm the validity of this assumption, we consider the porosity evolution given by Eq.~(\ref{eqPhiU}). 
The variable $\phi$ monotonically increases with the slip, reaching a limit of $\phi_\infty$. 
Therefore, we can expect that when $\phi_\infty$ is considerably smaller than unity, the dependence of COEs on $\phi$ can be neglected. 
Contrarily, when $\phi_\infty$ is not 
so small,
the dependence might not be negligible. 
Furthermore, it is important to assume that the porosity is not close to unity. 
Therefore, we utilize two specific values, $0.01$ and $0.1$, for $\phi_{\infty}$ in the numerical calculations below. 
The numerical calculations also make it possible to draw phase diagrams of the small and large slippages in the $p_0-\phi_0$ plane.

\section{Numerical Calculations} \label{secNC}

We numerically simulate the sequence of the repeating foreshocks and the mainshock 
by the direct integration of  Eqs.~(\ref{eqDPEL})--(\ref{eqeom}). 
The variables depend on space and time here, and those in the central part of the slip zone will be shown.  
We use the parameter values shown in Table 1 in Appendix A.
The quantities are normalized by 
$D_c \equiv C_1C_2 \equiv((1-\phi_\mathrm{ref})\rho_s C_s +\phi_\mathrm{ref} \rho_f C_f) / w_h ((b-\phi_\mathrm{ref}) \alpha_s +\phi_\mathrm{ref} \alpha_f) M_0 \mu_{\mathrm{slid}} 
\ [\mathrm{m}]$, $t_c \equiv D_c/\beta_v \ [\mathrm{s}]$, $\sigma_c =100 \ [\mathrm{MPa}]$, and $T_c =473.15 \ [\mathrm{K}]$.
We assume that $p_0=1$ and $\phi_0=0$ with the normalization at the onset of the numerical calculation. Additionally, we set $m=\sigma_c D_c t_c^2 \ [\mathrm{kg}]$ and $S=D_c^2 \ [\mathrm{m^2}]$, resulting in that the mass and the bottom area of the block are unity in the numerical calculations. 
The spring constant is fixed at $1 \times 10^6 [\mathrm{N m^{-1}}]$.
The values of $\phi_\infty=0.01$ and $0.1$ are investigated, as referenced in Sec.~\ref{secAT}. 
The Runge-Kutta method with the fourth-order accuracy is adopted. 
Note that performing a quantitatively exact comparison of the results with the natural fault slip behavior is difficult. 
Qualitative suggestions for the dynamic earthquake slip process are important, as mentioned in Sec.~\ref{secBM}. 

\begin{figure*}[tbp]
\centering
\includegraphics[width=12.5cm]{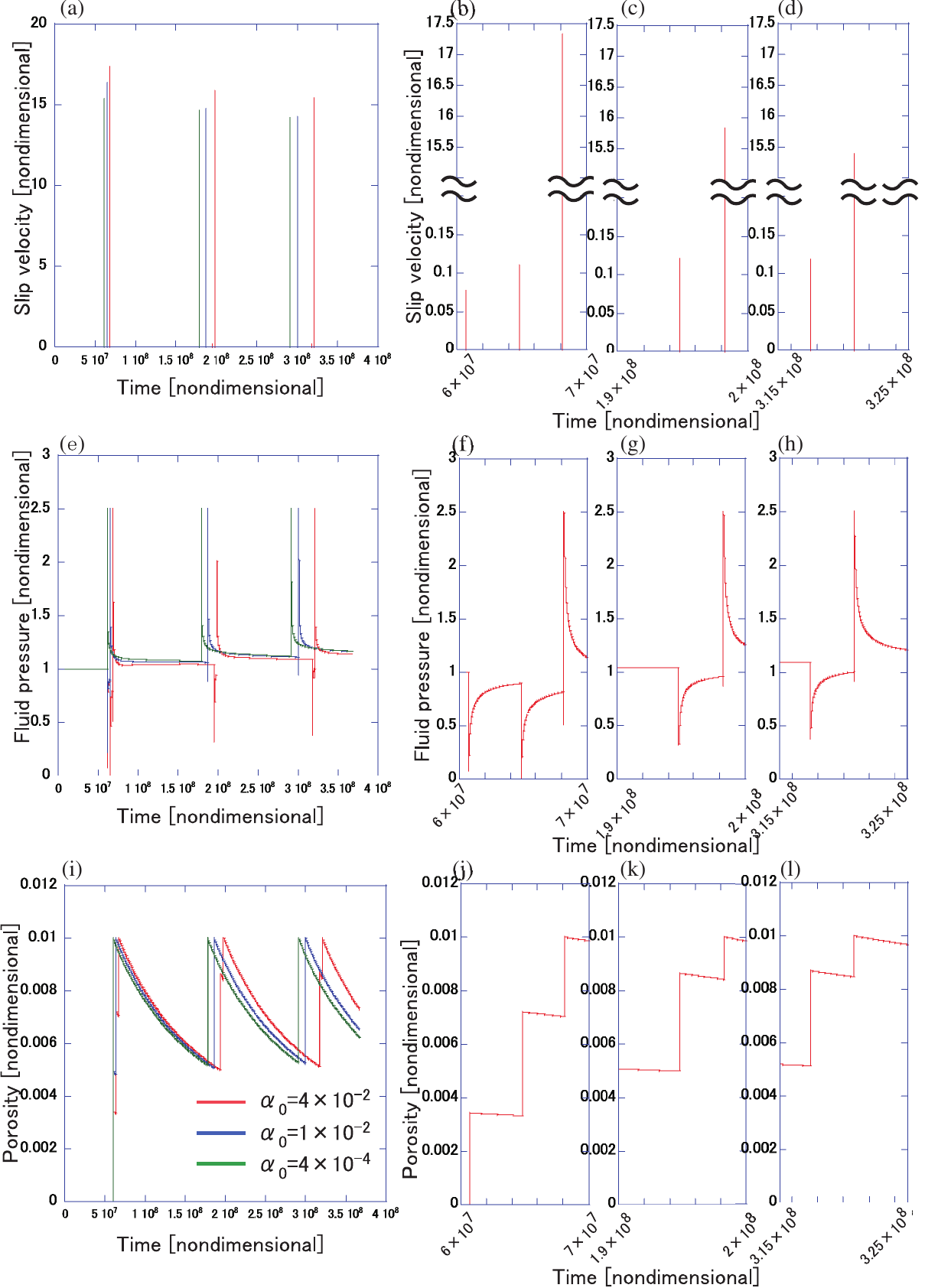}
\caption{Temporal evolutions of (a) $v$, (e) $p$, and (i) $\phi$ with $\phi_\infty=0.01$. For $\alpha_0=4 \times 10^{-2} [\mathrm{m}^{-1}]$, the temporal evolution of $v$ for three sequences of the mainshock and foreshocks are enlarged in (b), (c), and (d). The same regions are enlarged in (f), (g), and (h) for $p$ and in (j), (k), and (l) for $\phi$.}
\label{FigVPP1}
\end{figure*}

We first show the numerical results for $\phi_{\infty}=0.01$. The value of $\alpha_1$ is fixed at $10^{-5} \ [\mathrm{s}^{-1}]$, and three values of $\alpha_0$ are investigated: $\alpha_0=4 \times 10^{-2}, 1 \times 10^{-2}$ and $4 \times 10^{-4} \ [\mathrm{m}^{-1}]$. The temporal evolutions of $v$, $p$, and $\phi$ are shown in Fig.~\ref{FigVPP1}.  
We can observe that both small and large slippages occur with $\alpha_0=4 \times 10^{-2} \ [\mathrm{m}^{-1}]$ and $1 \times 10^{-2} \ [\mathrm{m}^{-1}]$, whereas only large slippages occur with $\alpha_0 =4 \times 10^{-4} \ [\mathrm{m}^{-1}]$. 
Two, one, and one small slippages, which are the foreshocks of the following mainshocks,  occur for the 1st, 2nd, and 3rd large slippages, which are the mainshocks, respectively, with $\alpha_0=4 \times 10^{-2} [\mathrm{m}^{-1}]$ . 
A single foreshock occurs for the 1st mainshock, whereas it does not occur for the 2nd and 3rd mainshocks with $\alpha_0=1 \times 10^{-2} [\mathrm{m}^{-1}]$. 
It turns out that the present model can simulate both the foreshock(s)-mainshock sequence and the mainshock repetition. 
The former sequence also repeats, i.e., mainshock-foreshosk transition also occurs.

\begin{figure}[tbp]
\centering
\includegraphics[width=8.5cm]{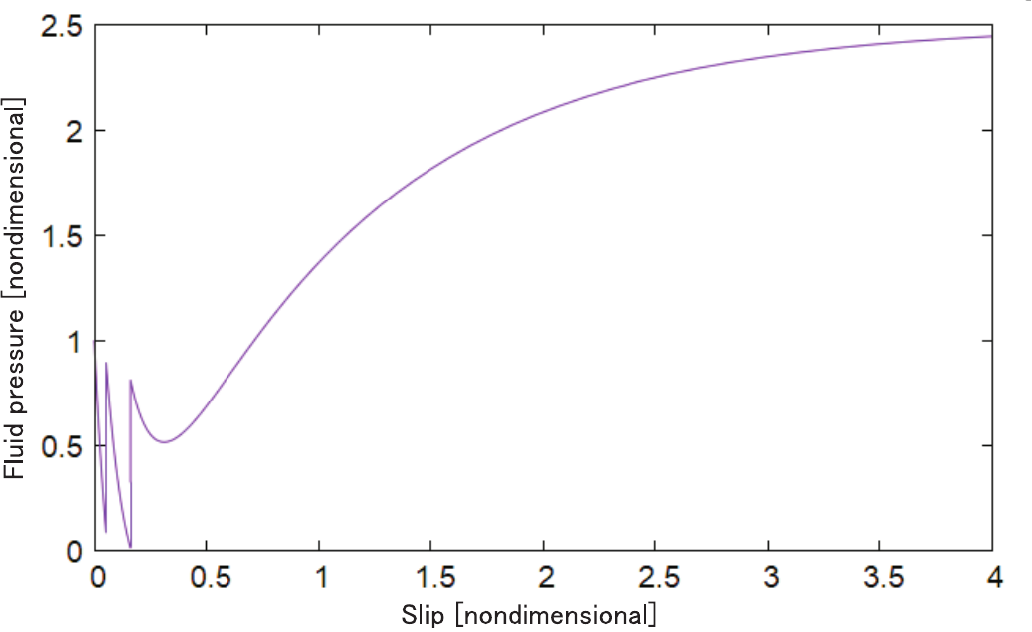}
\caption{The $u-p$ curve for (a) the 1st--3rd slippages with $\alpha_0=4 \  \times 10^{-2} [\mathrm{m}^{-1}] $ and $\phi_\infty=0.01$. }
\label{FigUP}
\end{figure}

The temporal change in the fluid pressure illustrated in Figs.~\ref{FigVPP1}e--h and \ref{FigUP} shows that it decreases for the small slippages. Notably, for the large slippages,  it  decreases at the onset of the slippage, after which it increases. This is because the pore-generation effect works at the onset of the slippage, whereas the effect becomes negligible during the large slippage. 
We can stipulate both thermal-pressurization and pore-generation effects as the reasons for the fluid-pressure change in the single slippage sequence. Further, we can confirm that $\phi$ does not exceed $\phi_\infty$ from Fig.~\ref{FigVPP1}i--l.

\begin{figure}[tbp]
\centering
\includegraphics[width=8.5cm]{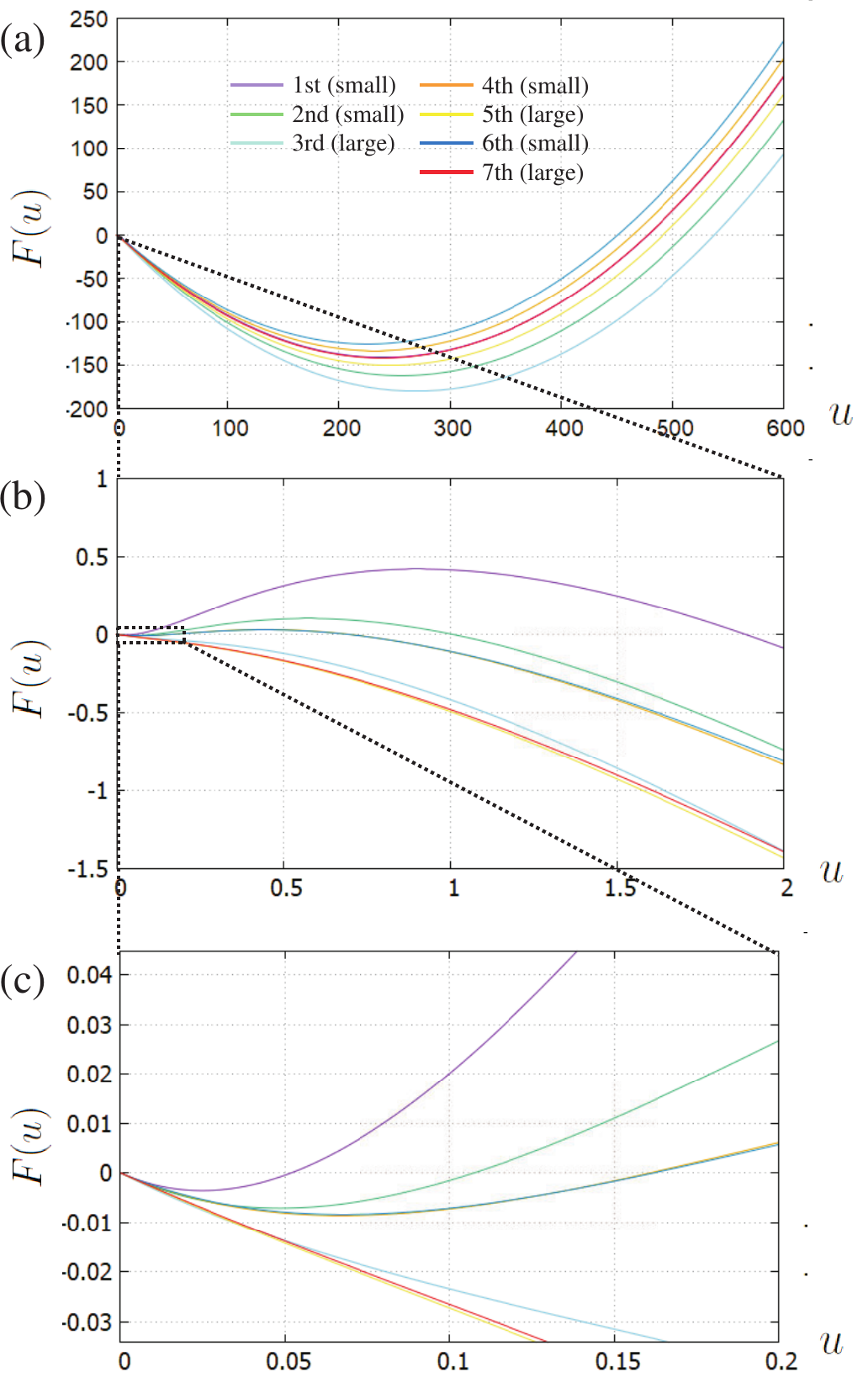}
\caption{Function $F(u)$ with $\alpha_0=4 \times 10^{-2} [\mathrm{m}^{-1}]$. The region near the origin in (a) is enlarged in (b), and the rectangular area near the origin in (b) is enlarged in (c).  The curves for the 4th (orange) and 6th (navy) almost overlie.}
\label{FigF1}
\end{figure}

Figure \ref{FigF1} shows the $u-F(u)$ curves for the seven slippages with $\alpha_0=4 \times 10^{-2}[\mathrm{m}^{-1}]$.
The figure indicates that the solutions of equation $F(u_f)=0$ for the small slippages are the small solutions, i.e., $u_{\rm small}$ in the Sec.~\ref{secAT}, because $F(u_2)>0$, whereas those for the large slippages are the large solutions, i.e., $u_{\rm large}$, because $F(u_2)<0$. 
This is consistent with the analytical treatment derived by neglecting the $\phi_0-$dependence of COEs, and we can conclude that the analytical treatment is valid with $\phi_\infty=0.01$.

\begin{figure}[tbp]
\centering
\includegraphics[width=8.5cm]{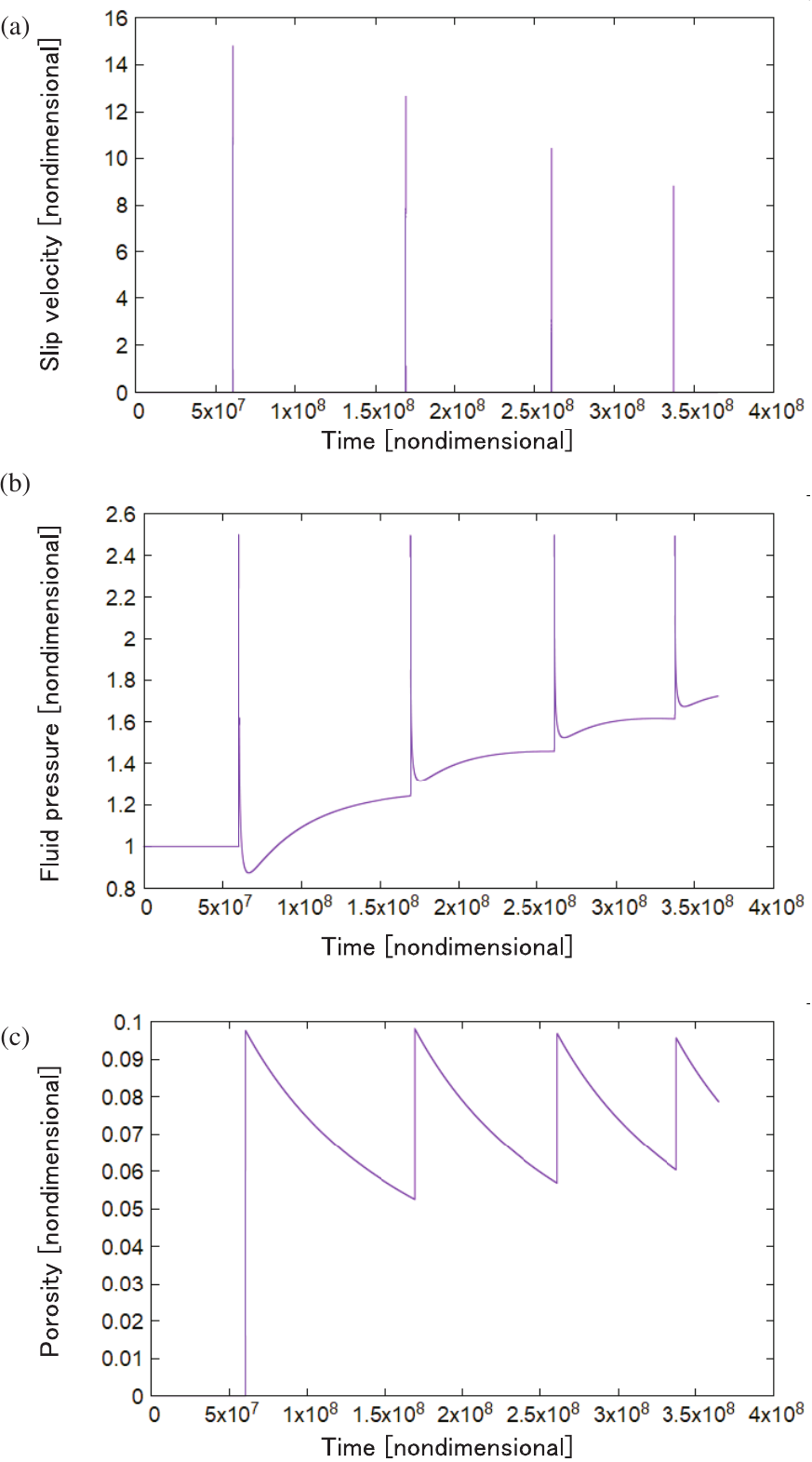}
\caption{Temporal evolutions of (a) $v$, (b) $p$, and (c) $\phi$ with $\phi_\infty=0.1$. }
\label{FigVPP2}
\end{figure}

Next, we examine the case of $\phi_\infty=0.1$. The value of  $\alpha_0$ is fixed at $4 \times 10^{-4} 
 \ [\mathrm{m}^{-1}]$. Figure \ref{FigVPP2} shows that only the large slippages are repeated. The fluid pressure increases with the slippages, and it gradually increases during the stick time (Fig.~\ref{FigVPP2}b). The porosity exhibits the cyclic-like behavior; however the change during the slippage decreases with the repetition of the slippages. With the $t \to \infty$ limit, $p$ and $\phi$ approach $-\sigma_n^0$ and $\phi_\infty$, respectively, and friction stress vanishes, as mentioned in Sec.~\ref{secIHFP}. 

\begin{figure}[tbp]
\centering
\includegraphics[width=8.5cm]{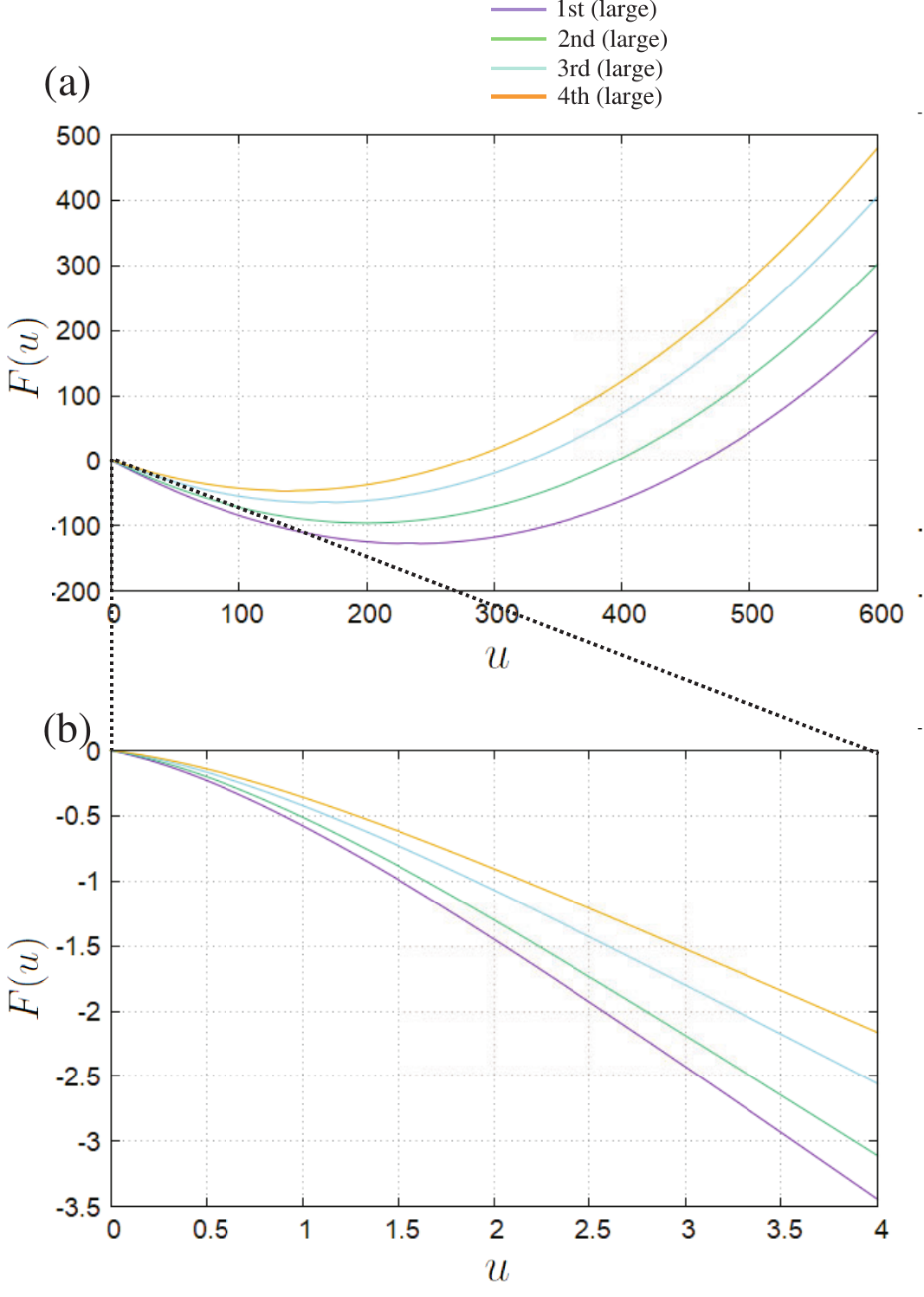}
\caption{Function $F(u)$. The region near the origin in (a) is enlarged in (b).  }
\label{FigF2}
\end{figure}

The behavior of the $u-F(u)$ curve for the four slippages with $\phi_\infty=0.1$ is shown in Fig.~\ref{FigF2}. 
For all slippages, equation $F(u_f)=0$ has only large solutions. The analytical prediction works well also for $\phi_\infty=0.1$. 
From the numerical calculation results, we can conclude that the analytical treatment is effective even if the COEs depend on $\phi$. 

The present model can simulate both the foreshock(s)-mainshock sequence and the mainshock repetition in the single model. 
Moreover, the former sequence also repeats, i.e., mainshock-foreshosk transition also occurs.
This is different from previous studies \cite{deS, Vas}. 
In particular, the transition from foreshock to mainshock and that from mainshock to foreshock cannot be realized based on the previous studies. 

We finally obtain the phase diagram of the small and large slippages, i.e., the phase diagram of the foreshocks and mainshock, in the $p_0-\phi_0$ plane, as follows.
From Eq.~(\ref{eqDiffCrit1}), equation $d F(u)/d u=0$ yields
\begin{equation}
e^{-\gamma' u_2}=\frac{k_p u_2}{\mu_{\mathrm{slid}} AS} +\frac{\mu_{\mathrm{stat}} (\sigma_n^0+p_0)}{\mu_{\mathrm{slid}}A} -\frac{\sigma_n^0 +p_0 -A}{A} e^{-\gamma u_2}.  \label{eqGU2}
\end{equation}
By definition, $d F(u)/d u|_{u=u_2}=0$ must be satisfied. Therefore, using Eqs. (\ref{eqCrit}) and (\ref{eqGU2}), we obtain
\begin{eqnarray}
\frac{1}{2} k_p u_2^2 +\left( \mu_{\mathrm{stat}} (\sigma_n^0+p_0) S +\frac{k_p}{\gamma'} \right) u_2 \nonumber \\
-\frac{\mu_{\mathrm{slid}}}{\gamma} (\sigma_n^0+p_0-A) S -\frac{\mu_{\mathrm{slid}}}{\gamma'}  AS \nonumber \\
+\frac{\mu_{\mathrm{stat}}}{\gamma'} (\sigma_n^0 +p_0) S \nonumber \\
+\mu_{\mathrm{slid}} (\sigma_n^0+p_0-A) S \left( \frac{1}{\gamma} -\frac{1}{\gamma'} \right) e^{-\gamma u_2}=0.  \label{eqU2}
\end{eqnarray}

When the magnitude of $\gamma u_2$ is greater than the order of unity, we can neglect the $\exp(-\gamma u_2)$ term.
Thus, we have a quadratic equation, which gives the relation among $u_2$, $p_0$, and $\phi_0$. 
By substituting the solution of this quadratic equation into Eq. (\ref{eqCrit}), we obtain $p_0$ and $\phi_0$, which are critical for inducing the phase transition, yielding
\begin{eqnarray}
F(u_2)&=&-\frac{k_p}{\gamma'} u_2 -\frac{\mu_{\mathrm{stat}}}{\gamma'} (\sigma_n^0+p_0)S +\frac{\mu_{\mathrm{slid}}}{\gamma'} e^{-\gamma' u_2} AS \nonumber \\
&\equiv& G(p_0, \phi_0) =0. \label{eqTC}
\end{eqnarray}

\begin{figure}[tbp]
\centering
\includegraphics[width=8.5cm]{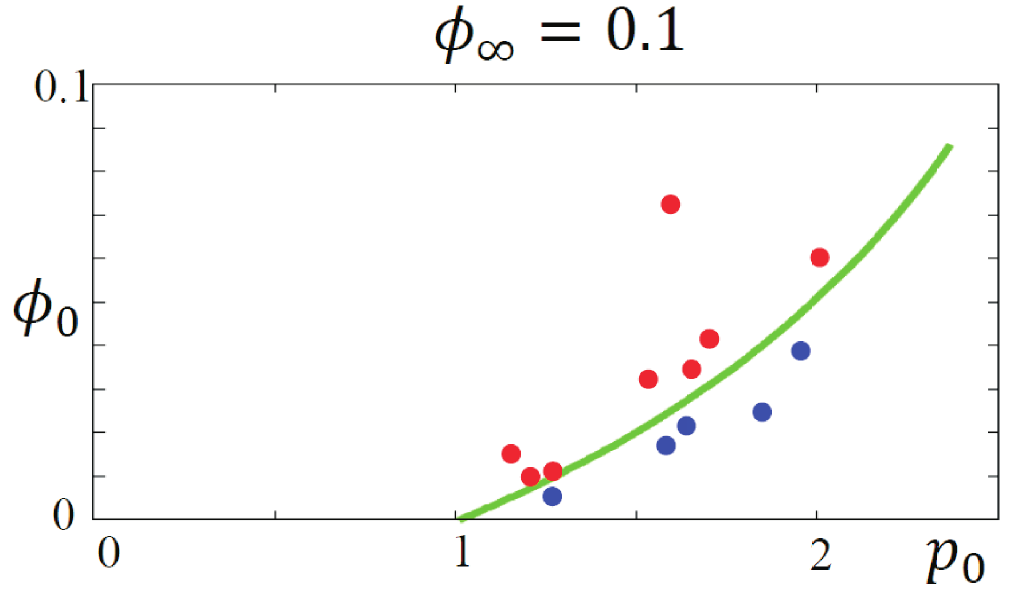}
\caption{Boundary between the small and large  slippages, denoted by the green line in the $p_0-\phi_0$ plane obtained from Eq.~(\ref{eqTC}).
The value of $\phi_\infty$ is 0.1.
The red and blue symbols show that $(p_0,\phi_0)$ cause large and small slippages, respectively, obtained by the direct integration of  Eqs.~(\ref{eqGP})--(\ref{eqGPhi}).
The approximation noted in the text is valid, and the phase boundary obtained analytically coincides well with that obtained numerically.
}
\label{Figpp2}
\end{figure}

The phase diagram thus obtained is shown in Figure \ref{Figpp2}, exhibiting the numerical results obtained using several sets of $(p_0, \phi_0)$ for $\phi_\infty=0.1$. 
The green line shows the phase boundary, $G(p_0, \phi_0)=0$, and the red and blue symbols show that $(p_0,\phi_0)$ cause large and small slippages, respectively. 
The analytical results coincide well with the numerical ones for $\phi_\infty=0.1$, and we can conclude that the approximation above is valid for this case.

Fast and slow earthquakes have recently attracted notable attention \cite{Oba, Kato12}. Small slippages can be considered slow earthquakes, whereas large slippages can be considered fast earthquakes. 
Figure \ref{Figpp2} indicates that high $p_0$ and/or low $\phi_0$ conditions tend to generate slow earthquakes. 
The observations  in the Nankai subduction zone in Japan show that high fluid-pressure areas generate slow earthquakes \cite{Par, She}. 
Moreover, such a high-fluid pressure is considered to be generated by the inhibition of fluid flow in low porosity areas \cite{Kata12}. 
These statements are consistent with the information in Fig.~\ref{Figpp2}.

\section{DISCUSSION AND CONCLUSIONS} \label{secDisCon}
In this paper it is shown that a transition from foreshocks to mainshocks occurs spontaneously in earthquake sequences analytically and numerically.
This transition is induced by the interaction among slip, frictional heating, fluid pressure, and porosity resulting from the fracture of fault rocks, and is described by the change in the number of solutions to the equation $F(u_f)=0$, which represents the energy balance 
before and after the slippage.
This transition can be regarded as a first-order phase transition. 
In particular, the values of $p$ and $\phi$ just before the dynamic slippage are found to play important roles in the transition. 
The phase boundary on the $p_0-\phi_0$ space has been obtained analytically via approximations, and the approximation was established to be effective with $\phi_\infty=0.1$.

Tremors can be simulated by introducing the fluid flow \cite{Suz10, Suz14}. 
Tremors can be regarded as events involving continuous low-frequency earthquakes (LFEs). 
The interval between the LFEs in the current study can be shortened by introducing fluid inflow, which reduces   the frictional stress, resulting in events akin to tremors.


The analytical treatment conducted here does not require a specific form of pore healing. 
The pore-healing and fluid-diffusion effects induce quantitative changes in the system behavior (i.e., the repeating times and the transition time change), and not qualitative changes.

\bmhead{Acknowledgements}

T. S. was supported by a Grant-in-Aid for Scientific Research (C), JP20K03771, JP23K03554, and JSPS KAKENHI Grant Number JP22H05309 in Transformative Research Areas (A) for FY 2021-2025 ``Science of Slow to Fast Earthquakes.''.  H. M. was supported by JSPS KAKENHI Grant Numbers JP20K03792, JP23K03252.  
This study was also supported by the Ministry of Education, Culture, Sports, Science and Technology (MEXT) of Japan, under its The Second Earthquake and Volcano Hazards Observation and Research Program (Earthquake and Volcano Hazard Reduction Research) and  by ERI JURP 2021-G-02, 2022-G-02, 2023-G-02, and 2024-G-05 in Earthquake Research Institute, the University of Tokyo. 

\section*{Declarations}

Data will be made available on reasonable request. Both T. S. and H. M. conceived of the presented idea. T. S. developed the theoretical formalism, performed the analytic calculations and performed the numerical simulations.  All authors provided critical feedback and helped shape the research, analysis and manuscript.

\appendix
\section{Parameter values}
The meaning of parameters and their values are summarized in Table  \ref{TabP}.
\begin{table*}{}
\caption{Properties, as well as their meanings and values. Values are based on Refs. \cite{Suz14, Suz17}. However, the values are solely employed for numerical computations.}
\begin{tabular}{lll}
\hline\hline
Properties & Physical Meanings & Values \\ \hline
$M_0=\left( \frac{b-\phi_t}{K_s} +\frac{\phi_t}{K_f} \right)^{-1}$ & Effective bulk modulus & $2.97 \times 10^4$ MPa \\
$\tilde{\alpha}=\left( (b-\phi_t) \alpha_s +\phi_t \alpha_f \right)$  & Effective thermal expansion coefficient & \\
$\tilde{C} = \left( (1-\phi_t) \rho_s C_s +\phi_t \rho_f C_f \right) $ & Effective heat capacity &  \\
$\tilde{\lambda}=\left( (1-\phi_t) \lambda_s +\phi_t \lambda_f \right) $ & Effective thermal diffusion constant & 0.2 \\
$b=1-\frac{K_v}{K_s} $ & & 0.2 \\
$C_s$ & Specific heat for the solid phase & $9.2 \times 10^2 \ \mathrm{J} \ \mathrm{kg}^{-1} \ \mathrm{K}^{-1}$ \\
$C_f$ & Specific heat for the fluid phase & $4.2 \times 10^3 \ \mathrm{J} \ \mathrm{kg}^{-1} \ \mathrm{K}^{-1}$ \\
$k$ & Permeability & $1 \times 10^{-20}$ $\mathrm{m^2}$ \\
$K_s$ & Bulk modulus of the solid phase & $3 \times 10^4$ MPa \\
$K_f$ & Bulk modulus of the fluid phase & $3.3 \times 10^3$ MPa \\
$K_v$ & Bulk modulus of the medium & $2.4 \times 10^4$ MPa \\
$\alpha_0$ & Parameter characterizing the porosity evolution for dynamic slips [see Eq. (\ref{eqDPhiEL})] & $4 \times 10^{-3}$ $\mathrm{m}^{-1}$ \\
$\alpha_1$ & Parameter characterizing the pore-healing effect during the stick time [see Eq. (\ref{eqSPhiEL})] & $1 \times 10^{-4} \mathrm{s}^{-1}$ \\
$\alpha_s$ & Thermal expansion coefficient of the solid phase & $1 \times 10^{-5}$ $\mathrm{K}^{-1}$ \\
$\alpha_f$ & Thermal expansion coefficient of the fluid phase & $2.1 \times 10^{-4}$ $\mathrm{K}^{-1}$ \\
$\beta_v=\sqrt{\frac{\mu}{(1-\phi_t)\rho_s +\phi_t \rho_f}}$ & Shear wave speed & $2.39 \times 10^3 \ \mathrm{m} \ \mathrm{s}^{-1}$ \\
$\eta$ & Fluid phase viscosity & $2.82 \times 10^{-4}$ Pa s \\
$\lambda_s$ & Thermal diffusion coefficient of the solid phase & $1 \mathrm{J} \mathrm{m^{-1}} \mathrm{K^{-1}} \mathrm{s^{-1}} $ \\
$\lambda_f$ & Thermal diffusion coefficient of the fluid phase & $6 \times 10^{-1} \mathrm{J} \mathrm{m^{-1}} \mathrm{K^{-1}} \mathrm{s^{-1}} $ \\
$\mu$ & Shear modulus of the medium & $1.44 \times 10^4$ MPa \\
$\mu_{\mathrm{slid}}$ & Sliding friction coefficient & 0.6 \\
$\mu_{\mathrm{stat}}$ & Static frictional coefficient & 0.8 \\
$\rho_s$ & Solid phase density & $2.7 \times 10^3 \ \mathrm{kg} \ \mathrm{m}^{-3}$ \\
$\rho_f$ & Fluid phase density & $1 \times 10^3 \ \mathrm{kg} \ \mathrm{m}^{-3}$ \\
$\sigma_n^0$ & Normal stress acting on the fault & $-2.5 \times 10^2$ MPa \\
$\sigma_{\mathrm{res}}=-\mu_{\mathrm{slid}} (\sigma_n^0 +p_0)$ & Residual frictional stress & -\footnotemark[1] \\
$\phi_t$ & Total porosity ($=\phi_\mathrm{ref}+\phi$) & -\footnotemark[1] \\ 
$\phi_\mathrm{ref}$ & Reference porosity before the slip & $0.1$ \\
\hline\hline
\footnotetext[1]{Function of time}
\end{tabular} \label{TabP}
\end{table*}

\renewcommand{\theequation}{A\arabic{equation}}
\setcounter{equation}{0}

%

\end{document}